\documentclass[A4paper,pre,twocolumn,showpacs,aps,floats,floatfix,superscriptaddress]{revtex4}
\pdfoutput=1

\usepackage{graphicx}
\usepackage{graphics}
\usepackage{subfigure}
\usepackage{epsfig}
\usepackage{amsmath,amssymb,bm}

\usepackage[colorlinks,linkcolor=blue,citecolor=blue]{hyperref}
\usepackage{dcolumn}
\usepackage{rotating}
\usepackage{multirow}
\usepackage{epsfig}
\usepackage{color}
\usepackage{float}
\usepackage{setspace}
\usepackage{natbib}
\usepackage{enumerate}

\begin{document}

\title{Spatial correlations in attribute communities}

\author{Federica Cerina} \affiliation{Department of Physics,
  University of Cagliari, Italy} \affiliation{Linkalab, Complex
  Systems Computational Laboratory, Cagliari, Italy} \author{Vincenzo
  De Leo} \affiliation{CRS4 Bioinformatica, c/o Parco Tecnologico
  Edificio 3 Loc. Piscina Manna, Pula, Italy} \affiliation{Linkalab,
  Complex Systems Computational Laboratory, Cagliari, Italy}
\author{Marc Barthelemy} \affiliation{Institut de Physique Th\'eorique
  (IPhT), CEA, CNRS-URA 2306, F-91191, Gif-sur-Yvette, France}
\author{Alessandro Chessa} \affiliation{Department of Physics,
  University of Cagliari, Italy} \affiliation{Linkalab, Complex
  Systems Computational Laboratory, Cagliari, Italy} \affiliation{Institute for Complex Systems, CNR UOS Department of Physics, University of Rome “Sapienza”, Rome, Italy}

\date{\today}

\begin{abstract}

  Community detection is an important tool for exploring and
  classifying the properties of large complex networks and should be
  of great help for spatial networks. Indeed, in addition to their
  location, nodes in spatial networks can have attributes such as the
  language for individuals, or any other socio-economical feature that
  we would like to identify in communities. We discuss in this paper a
  crucial aspect which was not considered in previous studies which is
  the possible existence of correlations between space and
  attributes. Introducing a simple toy model in which both space and
  node attributes are considered, we discuss the effect of
  space-attribute correlations on the results of various community
  detection methods proposed for spatial networks in this paper and in
  previous studies. When space is irrelevant, our model is equivalent
  to the stochastic block model which has been shown to display a
  detectability-non detectability transition. In the regime where
  space dominates the link formation process, most methods can fail to
  recover the communities, an effect which is particularly marked when
  space-attributes correlations are strong. In this latter case,
  community detection methods which remove the spatial component of
  the network can miss a large part of the community structure and can
  lead to incorrect results.

\end{abstract}


\maketitle 

\section{Introduction}

Many networks are embedded in real space and there is a cost
associated to the length of links. 
Examples of such spatial networks
can be found in infrastructures such as power grids, distribution and
logistic networks, transportation and mobility networks, and also in
computer science or biology with the Internet and neuronal networks
respectively (see for example the review
\cite{Barthelemy:2011}). Spatial constraints are so important in these
networks that one can expect a non-trivial spatial organization as
shown in various examples
\cite{Guimera:2005,Barrat:2005,Kaluza:2010,Expert:2010,Daraganova:2010,Chavez:2010,deMontis:2011,Ratti:2011,Brockmann:2011}.

In spatial networks, each node is described by its coordinates
(usually in a 2d space) but has in general other attributes. For
individuals, it can be any cultural or socio-economical parameter. For
infrastructure networks such as power grids, it can be the voltage at
the electric substations. In general, this attribute depends on space
and the resulting network displays entangled layers of parameters. An
important goal in the analysis of these networks is to disentangle
these different levels and to extract some mesoscopic information from
the spatial network structure. If one is interested in studying
effects beyond space \cite{Expert:2010}, one should have a
straightforward way to `subtract' it from the network, or in other
words, to disentangle space and the other attributes.

A natural tool for such a task is community detection which was used
for the characterization at a mesoscopic scale of the properties of
complex networks (see \cite{Fortunato:2010} for a review). A
(real-world) community can be naturally defined as a group of network
elements having the same attribute value such as language or age for
social networks, or the internet domain name for web pages. At a more
quantitative level, a community can be thought as a set of nodes more
densely linked with each other than with the rest of the network
\cite{Potter:2009}. Community detection procedures consist in finding
these groups of nodes in the network. Various methods were proposed so
far and we refer the interested reader to the review
\cite{Fortunato:2010}. In particular, the Newman-Girvan method
\cite{Newman:2004} which relies on the optimization of a quantity
called modularity is frequently used and despite its intrinsic limits
shown in \cite{Fortunato:2007}, it possesses the advantage of being
simple and relatively easy to implement.

Community detection can have several purposes in spatial networks
\cite{Guimera:2005,Kaluza:2010,Nussinov:2011b,Gregory:2011}, but probably the main
one is to disentangle these various aspects, including spatial
correlations of any type. In most cases
\cite{Guimera:2005,Kaluza:2010} communities are determined by the
geography only, which results from the simple fact that the most
important flows are among nodes in the same geographical regions. In
this sense, community detection in spatial networks offers a visual
representation of large exchange zones. This even suggests that
community detection might be an important tool in geography and in the
determination of new administrative or economical boundaries
\cite{deMontis:2011}.

In the general case, for a given network we don't know to what extent
the existence of a link between a pair of nodes is due to a specific
factor or to space only. The link could exist because of a strong
attribute affinity between the nodes, or in the other extreme case,
because they are close neighbors. In general, one could expect a
combination of these two effects. If we are interested in recovering
communities defined by an attribute (such as language for example)
from the network structure, we then have to consider various
assumptions such as the correlation between link formation, attribute
values and space. In order to understand the effect of the underlying
correlations, we can consider two extreme cases. When the links are
purely spatial and independent from the attributes, if we remove the
spatial component, we will observe random communities (obtained for a
random graph) which contain a random number of nodes with random
attributes. In this situation, community detection is unapplicable and
there is no way to recover attribute communities from the network
structure. The other extreme case is when the formation of a link
depends on the attributes only.  In this case, space is irrelevant and
any standard community detection method should give sensible results,
ie. communities made of nodes with the same attribute.

The important problem of interest here is thus the intermediate case
when the probability to have a link depends both on attributes and on
space. In this case we have to eliminate spatial effects in order to
recover the attribute structure. An important point in the discussion 
is then the existence of correlation between space and
attributes. The nature and existence of these correlations will govern
the way we will have to do community detection. In this paper, we
construct a simple artificial network model allowing us to investigate
the effect of these correlations on the results of the community
detection procedure. We will test various methods on this toy model.

\section{Materials and Methods}
\subsection{A BENCHMARK FOR SPATIAL NETWORKS WITH ATTRIBUTES}

In order to test these ideas and how community detection acts on
spatial networks, we define a simple model of spatial networks with
attributes. The attributes could be anything and we will restrict -
without loss of generality - to the simple binary case where the
attributes can have two possible values at each node. 
  We will introduce a simple model where nodes and their attributes
  are randomly distributed in space. In general, according to the
  various parameters of the model, the attributes can be delocalized
  in space or, on the contrary, be localized in some well-defined
  region. In some cases, some attribute community could emerge in
  space, but our target community structure will always be the
partition of the network in the two subgraphs composed of nodes with
the same attribute and we will test how various methods can recover
these two communities. In this respect the main focus
  of our work will be the disentanglement of the sole attribute
  network features beyond the spatial node arrangements. 

We construct the test (benchmark) network defining the vertex and edge 
properties in the following way.

{\bf{Vertex properties}}:

\begin{enumerate}
\item We generate points/nodes in the $2d$ space ($x - z$) in two
  spatial communities, say the North and the South, around the two
  centers $(x, z) = (0, + L)$ and $(x, z) = (0, - L)$ (see
  Fig.~\ref{fig:figure_1}). A simple way to do that is to
  generate points $i$ around the two centers according to the
  probability
  \begin{equation}
    p (x_i, z_i) \propto e^{- d_{ci} / \ell}
  \end{equation}
  where $d_{ci}$ is the euclidean distance between one of the centers
  $c$ and the node $i$ of coordinates $(x_i,z_i)$.

\begin{figure}[h]
 \centering
  \includegraphics[width=8cm]{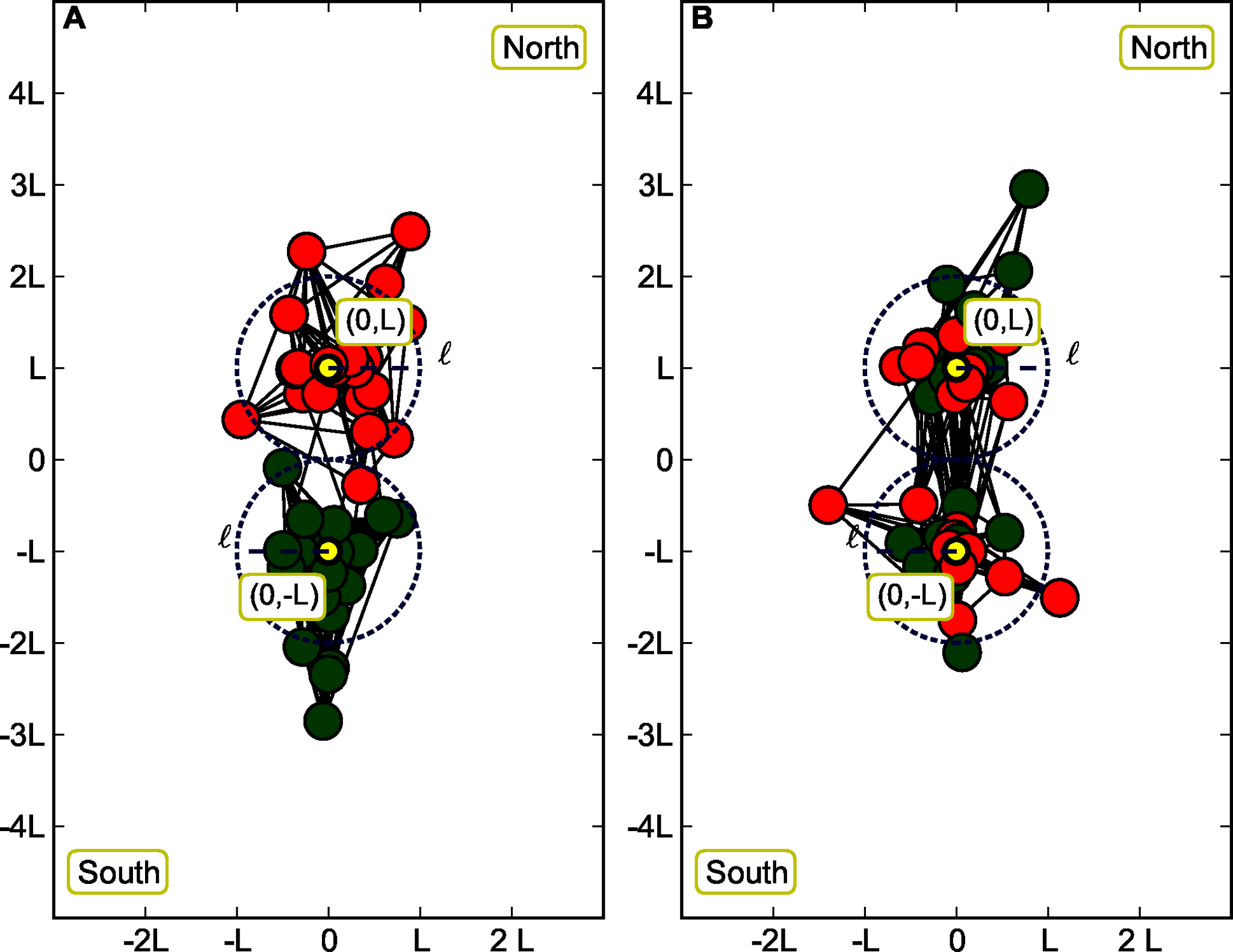} 
 \caption{The two spatial communities North and South are well
    separated having their average size $\ell=L$. In the
    A panel we present the case $\epsilon = 0$ where there is a
    perfect correlation between the space and the attributes (green
    and red colors). In the B panel, the uncorrelated case
    $\epsilon = 0.5$ is presented where the attribute colors are
    randomly distributed between the two segregated spatial
    communities (for the sake of clarity, only $40$ out of the $100$
    nodes used in our simulations are shown here{\color{black}, and $\beta=1.0$}).
    }
 \label{fig:figure_1}
\end{figure}

\item 
  We assign an attribute $S_i$ to each node $i$. In the following
  we will focus on the simplest case where this attribute can take only
  two values $S_i = \pm 1$ (which in this paper are the red and green
  colors).    
  A simple way to control correlations between attribute and
  space is to choose $S_i = + 1$ with probability $q$
  for $z > 0$ and $S_i=- 1$ with probability $1 - q$. 
  In order to tune the various cases we introduce the parameter $\epsilon$,
  with  $q=1-\epsilon$, that determines the mixing between space and attributes,
  ranging from $0.0$ to $0.5$. In the case  $\epsilon = 0.0$ space and attributes
  are strongly correlated, while for $\epsilon = 0.5$ space and attribute are totally uncorrelated.

  So the relevant parameters for the generation of network nodes are $\ell$ and  $\epsilon$.
  

{\bf{Edge properties}}:

\item We then construct the network: for each pair of nodes, we create
  a link between nodes $i$ and $j$ with probability $p_{link}(i, j)
  \propto e^{\beta S_i S_j - d_{ij} / \ell_0}$ where $\ell_0$ plays
  the role of the typical size of the spatial community (and where
  $d_{ij}$ is the euclidian distance between $i$ and $j$).
  {\color{black} It is worth observing that the parameter $l_0$ is the
    typical length of links when space dominates while $\ell$ is the
    typical spatial size of the northern and southern
    communities. Here the relevant edge parameters are $\beta$ and
    $\ell_0$, but in order to simplify the model and to focus on the
    efficiency of community detection methods, we choose
    $\ell=\ell_0$. This choice implies that when space dominates the
    link formation, the links cannot be much larger than the community
    size. In this case, the only spatial relevant parameter will be
    $\ell/L$ and we can fix $L$ to be equal to $1.0$ so that the
    spatial variability will be governed by $\ell$.  } We can rewrite
  the probability $p_{link}(i,j)$ as
\begin{equation}
  p_{link}(i, j)=\frac{1}{{\cal N}} e^{\beta (S_i S_j - d_{ij} / \ell \beta)} \label{eq:lbeta}
\end{equation}

 where ${\cal N}=\sum_{i<j}\exp{(\beta
    S_iS_j-d_{ij}/\ell)}$ is the normalization constant. As in the
  Erdos-Renyi random graph, the number of edges is a random variable
  with small fluctuations around its average. The number of nodes is
  thus fixed in each network but not the number of edges or the
  average degree, and this implies that we will have to average our
  observables over different realizations of the network.

When $\beta \ell$ is large, links are essentially between nodes with the same
attribute (irrespective of their distance) and if $\beta \ell$ is small then
space is the governing factor and links are essentially between neighboring
nodes.

\end{enumerate}

In this way the probability associated to a link depends on both space
and attribute, and the correlation between attributed and space can be
controlled. If the attribute is the same between two nodes the
probability to have a link will be reinforced, otherwise it will be
weakened, the interplay being controlled by the parameter $\beta$ .
Concerning the spatial factor, the closer the nodes and the larger the
probability associated to this link. 

The generation of attributes is an important point. We have two values
of the attribute only so that we need to generate attributes for only
half ($N / 2$) of the nodes.  So in the following we
  will study the specific case of an attribute community structure of
  equal size communities: half of the nodes has attribute $S_i=+1$ and the
  other half has $S_i=-1$.  We will investigate here two extreme
situations:
\begin{itemize}
  \item Attributes and space uncorrelated: this case is recovered by
    choosing $\epsilon=1/2$. 

  \item Attributes and space are strongly correlated. For this, we
    choose $\epsilon$ small. In this case, the spatial communities are
    also attribute communities.
 
\end{itemize}

\begin{figure}[h]
 \centering
  \includegraphics[width=8cm]{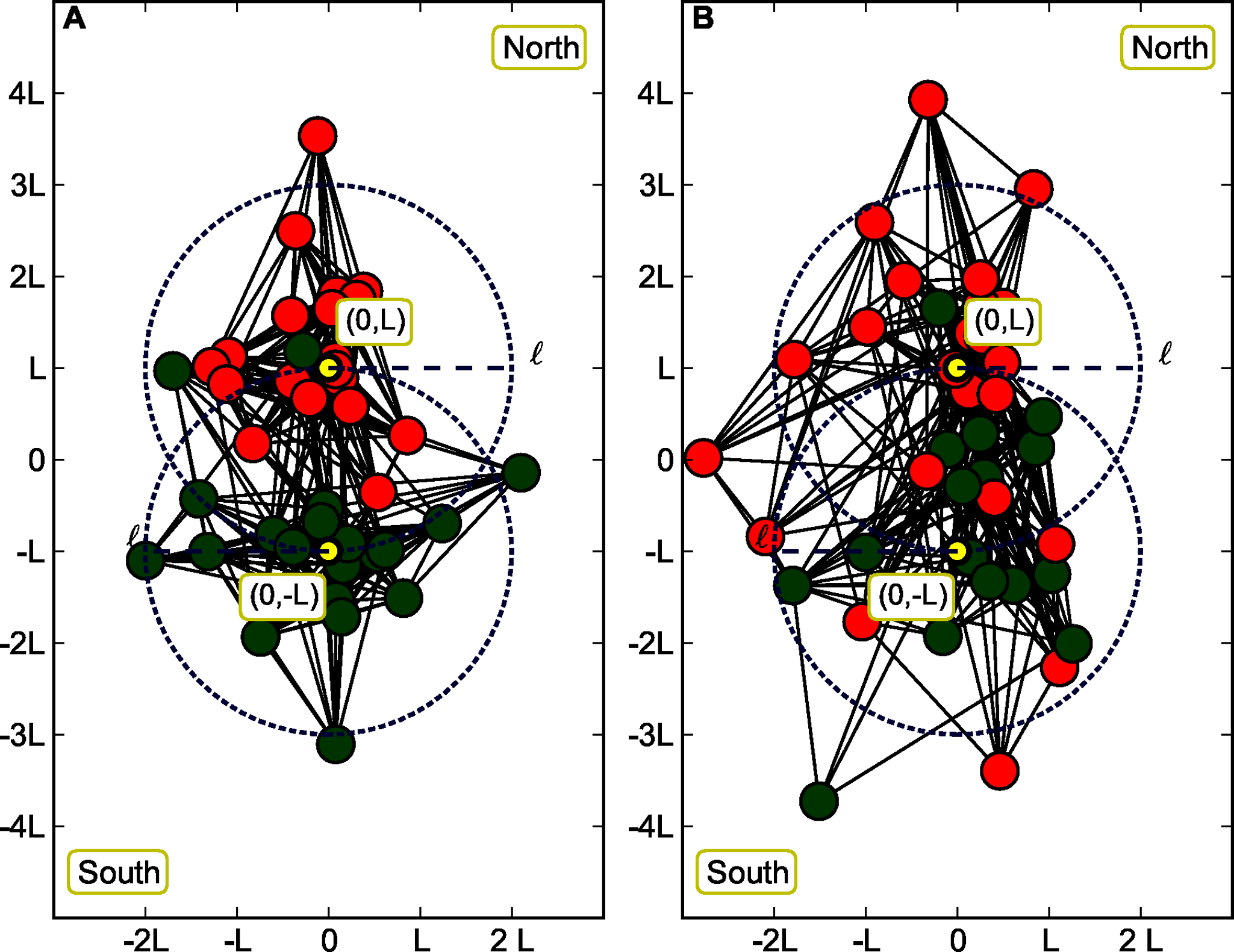} 
 \caption{The two communities North and South are mixing up each
    other with their average size $\ell$ approaching the value of
    $L$ (in this case $\ell=2L$). In the A panel, we display the case $\epsilon = 0.0$. Even
    if the spatial correlation is fading away the space-attribute
    correlation is still strong enough to display an attribute
    community. In the B panel, we show the extreme case $\epsilon
    = 0.5$ where the attributes are not correlated with space. In this
    case spatial mixing destroys the attribute community
    structure (for the sake of clarity, only $40$ out of the $100$
    nodes used in our simulations are shown here{\color{black}, and $\beta=1.0$}).
    }
 \label{fig:figure_2}
\end{figure}

Furthermore we can distinguish two different spatial arrangements for
the northern and southern communities. The first case corresponds to a
situation where the two communities are well separated with their
average size $\ell\le L$ and the spatial effects dominate the
community structure (see Fig.~\ref{fig:figure_1}). The second
situation corresponds to a larger value of the average community size
$\ell$ where the two communities start mixing up while $\ell$
approaches $L$ (see Fig.~\ref{fig:figure_2}).

There are many proposal in the literature for networks benchmarking
(see for example {\cite{Fortunato:2009}}), but this is -up to our
knowledge- the first one which takes into account the correlation
between space and node attributes.

\subsection{TESTING VARIOUS METHODS}

The interplay between space and attributes can lead to various situations that
need to be understood within the framework of community detection. Indeed we
have two main regimes $\beta \ell \gg 1$ and $\beta \ell \ll 1$ (see also Table \ref{table_1}):

\begin{center}
\begin{table*}[ht]
    \begin{tabular}{ | p{4.2cm} | p{5.3cm} | p{5.3cm} |}
    \hline
     Spatial correlation $\epsilon$ & $\beta\ell \ll 1$: Space is the
     governing factor & $\beta\ell \gg 1$: The spatial component of the
     links is irrelevant \\ \hline
	Spatially correlated: 
	
	($\epsilon=0.0$) & \begin{itemize}
	\item Links are between neighboring nodes but spatial communities correspond to the attribute ones.
	\item Any regular community detection will work.
	\end{itemize}    
	& \begin{itemize}
	\item Links are between nodes with the same attribute.
	\item Any community detection method should work.
	\end{itemize}
	\\ \hline
	Spatially uncorrelated: 
	
	($\epsilon=0.5$) & \begin{itemize}
	\item Links are between neighboring nodes but the attributes are anywhere in space.
	\item  It is necessary to `remove' space in order to uncover the attribute communities. 	
	\end{itemize} 
	& \begin{itemize}
	\item Links are between nodes with the same attribute.
	\item Any community detection method should work.
	\end{itemize}
	\\ \hline
    \end{tabular}
    \caption{ The table gives an account of the behaviour of the model in the regimes $\beta\ell \ll 1$ and $\beta\ell \ll 1$ both in the correlated ($\epsilon = 0.0$) and uncorrelated ($\epsilon = 0.5$) case. }
  	 \label{table_1}
  \end{table*}
\end{center}

\begin{enumerate}[(a)]
  \item $\beta \ell \gg 1$. In this case, the spatial component of the links
  becomes irrelevant (see Eq. \ref{eq:lbeta}) and for a given value of $\beta$ the
  community structure due to the node attributes will emerge, independently
  from the correlation between space and attributes. In this regime any
  community detection method should work.
  \item $\beta \ell \ll 1$. Here we have two subcases depending con the correlation between space and attributes:
  \begin{itemize} 
  \item ($\epsilon=0.0$) Space and attributes are \underline{correlated}: any regular community detection will work and moreover if
  you carefully remove the spatial effect the attribute community structure will be recovered.
  \item ($\epsilon=0.5$) Space and attributes are \underline{uncorrelated}: in this case the links are between neighboring
  nodes but the attributes are anywhere in space. Standard community detection methods
  won't work and it is then necessary to `remove' space in order to uncover the attribute communities. 
  \end{itemize}
\end{enumerate}

The general assumption of our model is to what extent it is possible to detect communities even if there is a spatial influence. Without space the initial situation is clear: we have two communities by construction and the probability of two nodes to be connected is related to the attribute similarities. Nodes with S=+1  tend mainly to connect to each other and the same for the S=-1 nodes. If we then put nodes in space and enhance the connection probability due to the proximity of nodes, it is not clear if a regular community detection method is able to detect the original two communities structure.
We thus see that correlations between space and attributes can be
misleading and any community detection method for spatial networks
should take into account this problem. There are now many community
detection methods \cite{Fortunato:2010} and in the following we will
use modularity optimization introduced by Newman and Girvan
{\cite{Newman:2004}}. This method suffers from various problems, the
most important being the existence of a resolution limit
\cite{Fortunato:2007} which prevent it to detect smaller modules, but
it is simple enough to implement. In addition, our point here is to
understand the effect of space-attributes correlations on community
detection and not to compare various methods.  In the following we
will thus essentially probe the Newman-Girvan method and variants
proposed here and in \cite{Expert:2010} for cases where the space and
attribute have different degrees of correlation.

The modularity function which needs to be optimized is defined as \cite{Newman:2004}:
\begin{equation}
  \textit{Q} = \frac{1}{2 \textit{m}} \sum_{ij} (A_{ij} - P_{ij}) \delta
  (C_i, C_j)
\end{equation}
where the sum is over all the node pairs, $A$ is the adjacency matrix,
\textit{m} is the total number of edges and $P_{ij}$ is the expected
number of edges between the vertices \textit{i} e \textit{j} for a
given null model. The $\delta$ function will result in a null
contribution for couples of vertices not belonging to the same
community ($C_i \neq C_j$). For an unweighted network, one can choose
$P_{ij} = \frac{k_i k_j}{2 m}$ which amounts to take as a null model a
random network with the same degree sequence as the original
network. In order to introduce explicitly space, the idea is to change
the null model defined by $P_{ij}$ and to compare the actual network
with this null model. Recently, such a proposal was made in
\cite{Expert:2010} where the quantity $P_{ij}$ is directly obtained
from the data describing the network. More precisely, Expert et
al. \cite{Expert:2010} used the following form
\begin{equation}
  P^{Data}_{ij} = N_i N_j f(d_{ij})
\end{equation}
where $N_i$ is related to the importance of the node $i$ (such as the
population for example). This form is reminiscent of the gravitional
model for traffic flows (see for example \cite{Erlander:1990}) where
flows are proportional to the product of populations and decrease
with distance. In \cite{Expert:2010}, the authors proposed to estimate
the unknown function $f$ directly from the empirical data by
\begin{equation}
f(d)=\frac{\sum_{i, j|d_{ij} = d} A_{ij}}{\sum_{i,j|d_{ij} = d} N_i N_j} \label{eq:expert}
\end{equation}
which can be seen as the probability to have two nodes connected at a
distance $d$.  Note that there is a binning procedure
  hidden in Eq. (\ref{eq:expert}).  The usual way to proceed in these
  cases consists in introducing a discretization of the space in bins
  that capture classes of distances.  Following \cite{Expert:2010}, we
  performed a binning of distances selecting the best value for the
  number of bins after a detailed stability study of the distributions
  obtained from the data. 

Expert et al. \cite{Expert:2010} applied this method to the specific
case of the phone network in Belgium, and try to reconstruct
linguistic communities (Flemish and French) beyond individuals spatial
location. This choice is probably the best one if there are no
correlations between the attribute under study (in their case the
linguistic membership of the people calling each other) and space. In
this specific case, extracting the node spatial dependencies from the
actual link distribution present in the network data is the most
effective way to subtract the spatial component. Otherwise if there are any
correlations between space and node attributes, the data contain in
an unknown proportion the two informations (space and attribute) and
their method needs to be reformulated. One possible way to do this is to explicitly
guess a spatial dependency of the link distribution and to put it as
an independent factor in the optimization function definition. In
order to be able to deal with the correlated case and to remove
spatial effect only, we thus
propose the following explicit function of space for $P_{ij}$
\begin{equation}
  P^{Spatial}_{ij} = \frac{1}{Z}k_i k_j g(d_{ij})
\label{eq:spatial}
\end{equation}
where $Z$ is the normalization constant, $k_i$ the degree of the node
$i$, $d_{ij}$ the euclidean distance between node $i$ and node
$j$. The function $g(d)$ is a decreasing function of distance and its
role is to remove the spatial effect. A simple choice is
\begin{equation}
g(d)=e^{- d/{\overline{\ell}}}
\end{equation}
where $\overline{\ell}$ is the average distance between nodes in the
network. Of course $\overline{\ell}$ is a rough approximation of the
real $\ell$ value, but we will see in the following that it is enough
to capture the essence of the spatial signature of the network.

We now need a method to compare the community structure obtained with
the modularity optimization and the expected one for the attribute
membership.  Many proposals have been introduced
{\cite{Danon:2005,Campello:2007,Karrer:2008}}, and we decided to use
here the \textit{Jaccard Index} {\cite{Jain:1988,Halkidi:2001}}. This
index is an extension of the Rand index {\cite{Rand:1971}}, and is
considered to be one of the most robust measure for the clustering and
classification assessment of graphs {\cite{Denoeud:2005}}. If $C$ is
the partition to be evaluated and $C'$ the reference one the
definition is as follows
\begin{equation}
  J_I = \frac{a}{a + b + c}
\end{equation}
{\color{black} where $a$ is the number of vertices pairs that are in the same
community for both $C$ and $C'$, $b$ is the number of pairs that are in different 
communities in $C$ but in the same one in $C'$  and
finally $c$ is the number of vertices pairs that are in the same
community in $C$ but not in $C'$ (or conversely).  
}
This quantity $J_I$
is in the interval $[0,1]$ and the closer to one, the better the
agreement between the two partitions. For $J_I = 1$ there is a perfect
match between the two community structures. In our case, it would mean
that the attribute communities are exactly detected. For values of
$J_I$ less than $1$ the discrepancy can depend both on the size of the
partitions in the community structure and/or the number of them and in
this respect the \textit{Jaccard Index} is a good method to compare a
very heterogeneous range of community structures.

\begin{figure}[h]
 \centering
  \includegraphics[width=8cm]{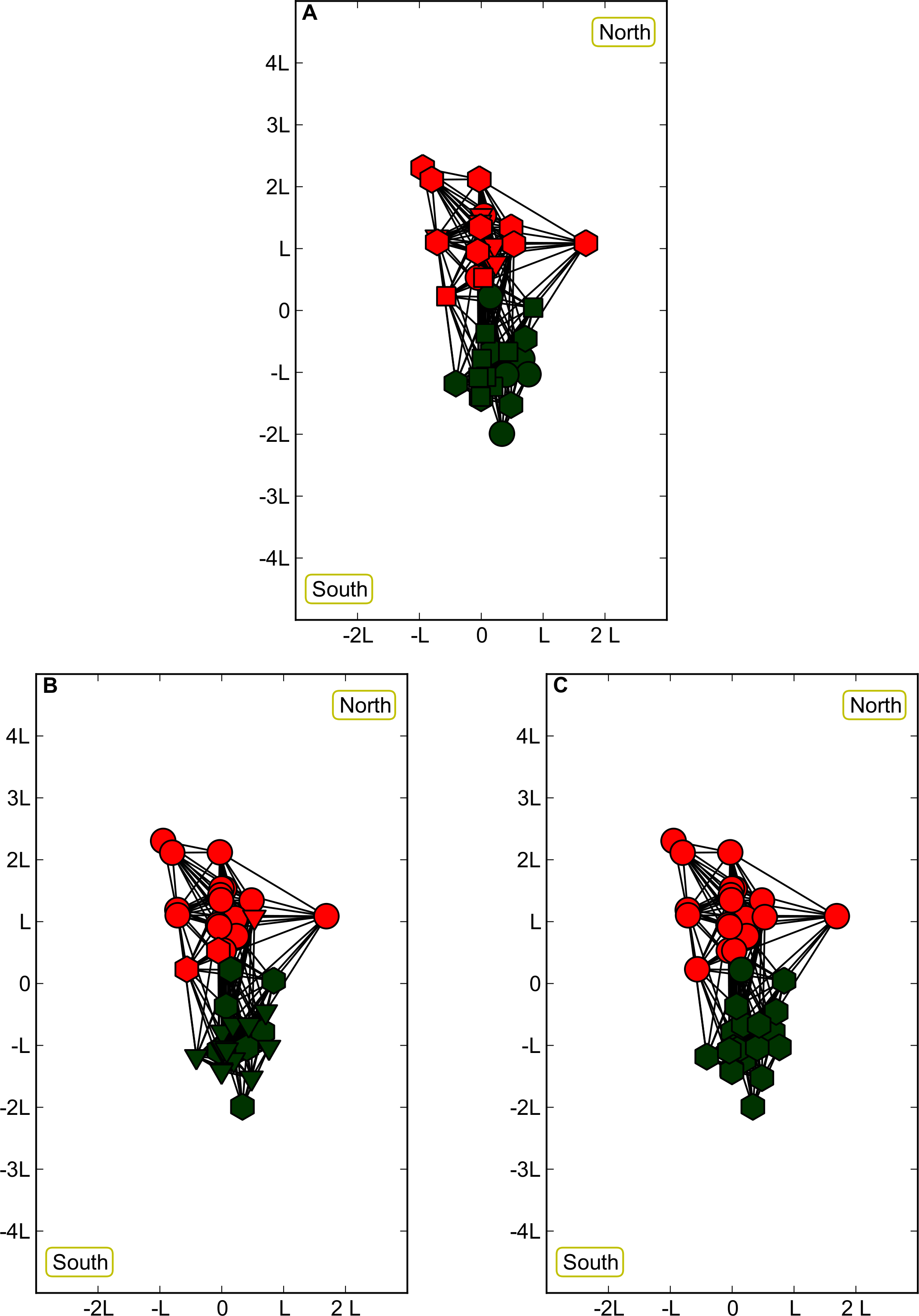} 
 \caption{Three spatial network configurations are presented for the
    constant value $\beta \ell=0.2$ and the correlated case
    $\epsilon=0.0$ with $\ell=1.0$ and $L=1.0$. The color (red and
    green) are the attributes, while the geometrical shapes represent
    the community memberships found with the various community
    detection procedure discussed in this paper. In the A panel, we
    present the case $J_I = 0.232$, obtained with the Data method. Due
    to the low $J_I$ value four communities are present (instead of
    the two associated with the attributes in red and green colors)
    and they are also mixed up between the south and the north spatial
    regions.  In the B panel we show the $J_I = 0.579$ case obtained
    with the Spatial method. Three communities are present and in the
    northern part there is a prevalence of circles while in the
    southern of triangles.  The C panel displays the case $J_I =
    0.903$ obtained with the Newman-Girvan formulation and the
    attribute community structure is almost completely recovered.
    }
 \label{fig:figure_3}
\end{figure}

In order to get a more intuitive picture of the Jaccard index, we show
three different cases in Fig.~\ref{fig:figure_3} for the same value
$\beta \ell=0.2$ (and in the case $\epsilon=0.0$, $\ell=1.0$ and
$L=1.0$) but with different values of $J_I$. The first case
corresponds to a relatively small value $J_I = 0.232$ (obtained with
the 'Data' method of \cite{Expert:2010},  where the binning is done as in their paper, which shows a partition in four communities (instead of 
the two associated with the attributes in red and green colors). For
intermediate values such as $J_I = 0.579$ (obtained with our 'Spatial'
method) the communities reduce to three with a prevalence of circles in the nothern part
and triangles in the southern (see B panel in Fig.~\ref{fig:figure_3}). 
The last case (obtained with the original Newman-Girvan formulation) corresponds to
a value $J_I =0.903$, that almost recovers the attribute community structure.

Finally, in order to have a baseline value we also computed the
average Jaccard for a completely random partition for $N=100$ nodes
and we obtain the value $J_I=0.08\pm 0.05$.

\section{Results}

The goal of this spatial community detection is to substract the spatial component and to recover the (two) attribute communities. We thus have three community detection methods: the original
Newman-Girvan method, the `Data' method proposed in
\cite{Expert:2010}, and our `Spatial' method defined by the null model of
Eq. (\ref{eq:spatial}) and, in order to understand their limits , we will test them against the benchmark network introduced
above. 

We will now see how these three different methods perform in the two
extreme cases of attribute correlated ($\epsilon = 0$) and
uncorrelated ($\epsilon = 0.5$) with space, both varying the size of the spatial communities $\ell$
and the attribute linkage strength $\beta$. 
The size of the test network is $N = 100$
nodes and the number of links depends on the probability previously
defined (Eq.~\ref{eq:lbeta}). We generated $100$ network
realizations for each set of parameters ($\beta$, $\ell$, $\epsilon$
and $L=1$). For each point of the simulation curve the error bars are
the standard deviation for $100$ modularity measures. To optimize the
modularity we used the Louvain method {\cite{Blondel:2008}}.

\begin{figure}[h]
 \centering
  \includegraphics[width=8cm]{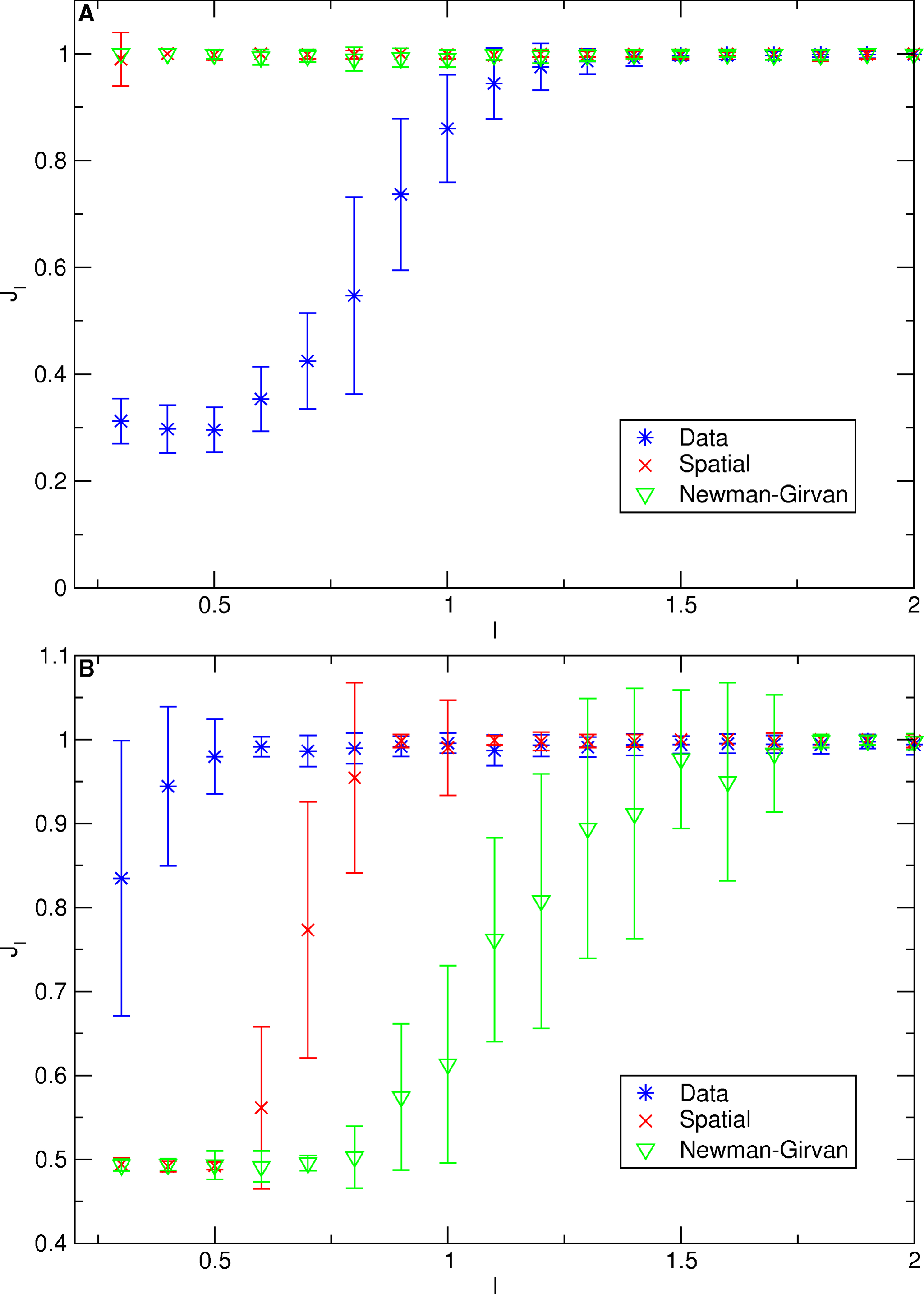} 
 \caption{The community structure obtained for various values of
  {\color{black}$\ell$} with fixed $\beta=1.0$. Each point represents
  the average Jaccard index for $100$ network community detection and
  the error bar is its standard deviation. The correlated case
  $\epsilon=0$ is shown on the A panel, and on the B panel we show the
  uncorrelated case $\epsilon=0.5$.  In A for  the
    regime $\beta \ell \ll 1$ both the Newman-Girvan and the
  'Spatial' method formulations give the right attribute community
  structure corresponding to the Jaccard index $J_I = 1.0$. For
  the regime $\beta \ell \gg 1$ all the three
  formulations work well since the links due to the attribute
  similarity are strong enough to preserve the community structure
  irrespectively from the node's location. In the uncorrelated case (B
  panel), the Data based formulation performs better respect to the
  Spatial formulation, since it extracts correctly the spatial
  information, directly from the data. In any case both spatial
  methods reach the right attribute community structure at almost the
  same value for {\color{black}$\ell \simeq 1.0$}. The Newman-Girvan
  standard formulation instead fails to detect the correct result up
  to values of {\color{black}$\ell \simeq 1.8$}. Note
    that in the x-axis we considered only values equal or above $0.3$
    since we verified that below this value the model generates
    disconnected networks.
    }
 \label{fig:figure_4}
\end{figure}

The behavior of the model depends on both parameters $\beta$ and
$\ell$ and we will first show the case with fixed attribute strength
$\beta$. We show on the A panel of figure \ref{fig:figure_4} the
correlated case ($\epsilon =0$) with a fixed $\beta=1.0$.

In this case, for $\beta \ell \gg 1$, all the three methods work well,
as expected and we obtain a perfect match ($J_I = 1$) between the
community structure resulting from the modularity optimization and the
attribute communities. Space is not relevant in this regime and links
exist essentially among nodes with the same attribute. For $\beta \ell
\ll 1$ both the Newman-Girvan modularity and the 'Spatial' method give
the correct result.  The latter actually subtract only the spatial
dependency while the the 'Data' method mixes the space effect with the
correlated attribute feature, resulting in a wrong community
detection.  The 'Data' method, for a sufficiently large value of
{\color{black}$\ell$} will approach anyway the correct $J_I=1.0$ value.

In the uncorrelated case (Fig.~\ref{fig:figure_4}, B panel) and for
a low values of $\beta \ell$, the Newman-Girvan modularity is not able
to detect the right attribute communities, since the attribute
correlation is not strong enough to group together the nodes of
similar type. Instead the other two methods perform better
in getting the attribute communities since they are able to correctly
eliminate the effect of space and recover the attribute community
structure, even for a small attribute correlation. The formulation based
on Data performs even better since it eliminates the effect of
space almost pointwise, but in any case the correct result of $J_I =
1$ is reached almost at the same value {\color{black}$\ell \simeq 1.0$} for
both spatial methods.

\begin{figure}[h]
 \centering
  \includegraphics[width=8cm]{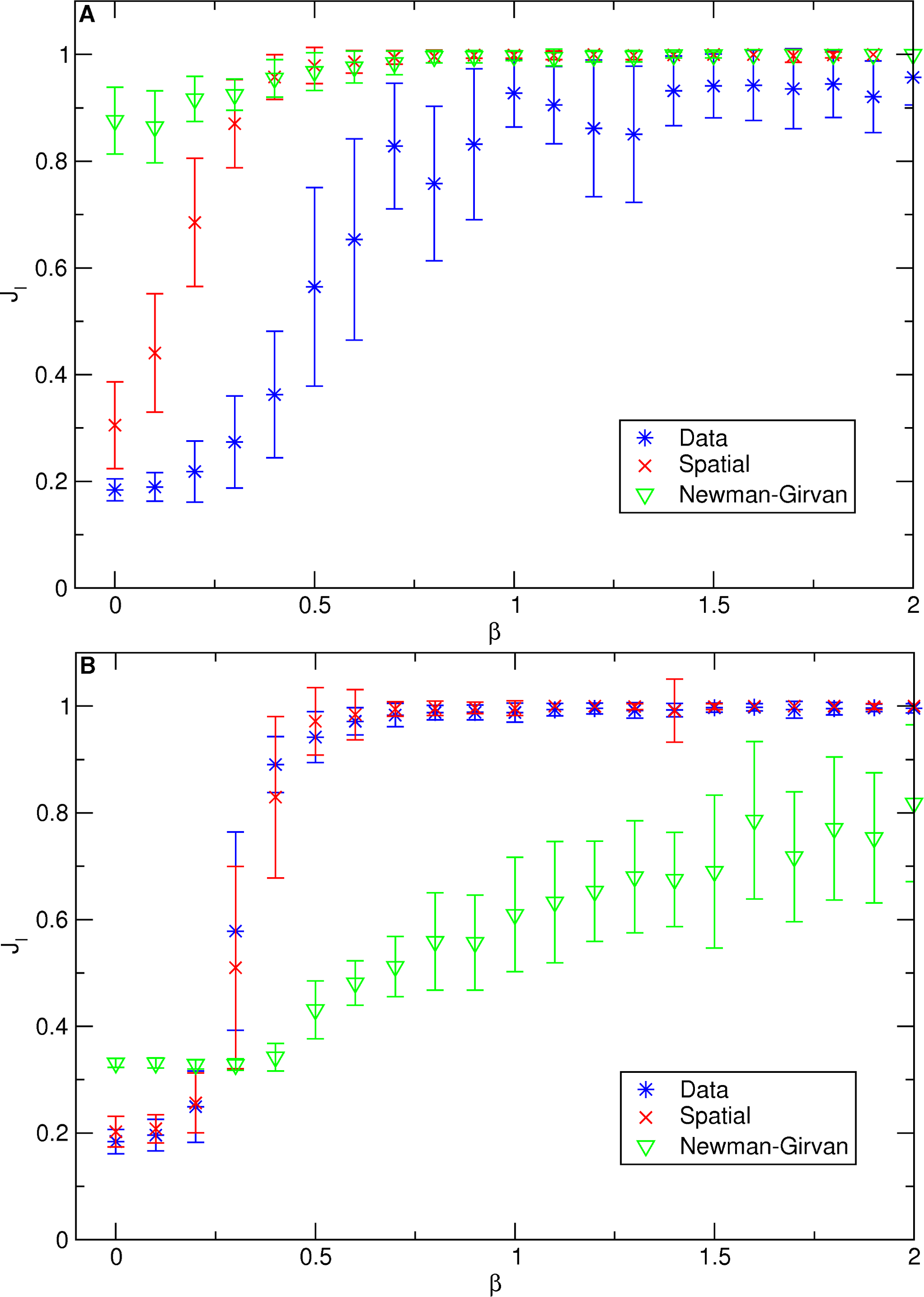} 
 \caption{The community structure obtained for various values of {\color{black}$\beta$} 
with fixed community size $\ell=1.0$. Each point represents
  the average Jaccard index for $100$ network community detection and
  the error bar is its standard deviation. The correlated case
  $\epsilon=0$ is shown on the A panel, and on the B panel we show the
  uncorrelated case $\epsilon=0.5$. In the uncorrelated case the
  'Data' method fails in detecting the attribute community structure
  for all the $\beta\ell$ regimes present in the figure, while the
  other two methods start working at {\color{black}$\beta=0.8$}. In the
  uncorrelated case the Newman-Girvan method is not able to detect the
  attribute community structure, while the spatial methods perform
  similarly better approaching the correct $J_I=1.0$ value around
  {\color{black}$\beta=0.8$}.
    }
 \label{fig:figure_5}
\end{figure}

In Figure~\ref{fig:figure_5} we show the results for the case of a
fixed community size ($\ell=1.0$) but where we vary the attribute
strength $\beta$. In the A panel the correlated case is presented
($\epsilon=0$). As expected the `Data' method for low values of {\color{black}$\beta$}
has problems in detecting the attribute community structure and
only for high attribute strengths ($\beta$) it starts to correctly
detect the target communities. In the uncorrelated case, where the
space is irrelevant, the standard Newman-Girvan formulation fails,
while the two spatial methods performs similarly better
(Fig.~\ref{fig:figure_5}).

{\color{black} In order to summarize these results we show
  in Table \ref{table_2} the only relevant regime (b) previously
  defined, $\beta \ell \ll 1$ (the (a) regime $\beta \ell \gg 1$ is
  trivial as we can verify in Figs \ref{fig:figure_4} and
  \ref{fig:figure_5}) for all the parameters of interest ($\epsilon$,
  $\ell$ and $\beta$) and for the three community detection methods. From
  this Table, it clearly emerges that the Spatial method is a very good
  interplay in all situations, while to get the best performances one
  has to choose the suitable method for any specific case.  }

\begin{center}
  \begin{table}
    \begin{tabular}{|cc|c|c|c|c|}
    \hline
     Spatial correlation $\epsilon$ &  & Newman-Girvan & Data & Spatial \\ \hline
    \multicolumn{1}{|c|}{\multirow{2}{*}{$0.0$ (correlated)}} &
	\multicolumn{1}{|c|}{$\ell$} & VG & B & VG   \\ \cline{2-5}
	\multicolumn{1}{|c|}{}                        &
	\multicolumn{1}{|c|}{$\beta$} & VG & B & G    \\ \cline{1-5}
	\multicolumn{1}{|c|}{\multirow{2}{*}{ $0.5$ (uncorrelated)}} &
	\multicolumn{1}{|c|}{$\ell$} & B & VG & G    \\ \cline{2-5}
	\multicolumn{1}{|c|}{}                        &
	\multicolumn{1}{|c|}{$\beta$} & B & G & G   \\ \cline{1-5}
    \hline
    \end{tabular}
    \caption{The table summarizes the performances, as can be extracted from Figs \ref{fig:figure_4} and \ref{fig:figure_5}, of the three methods (Newman-Girvan, Data and Spatial) in the only non trivial regime $\beta\ell \ll 1$, both in the correlated ($\epsilon = 0.0$) and uncorrelated ($\epsilon = 0.5$) case. Since in the plots we vary both $\ell$ and $\beta$, we distinguish here these two cases. In order to be able to compare this results we classified them according to the following criteria: \textbf{B}, \textbf{G} and \textbf{VG} that stand for \textbf{Bad}, \textbf{Good} and \textbf{Very Good}. We assign VG when there is a very good agreement with the target attribute community structure ($J_I$ very close to $1$), G when the behavior is rapidly approaching the correct result even for low/medium values of the parameters $\ell$ and $\beta$, and finally B when it completely fails to recover the right community structure.}
  \label{table_2}
 \end{table} 
\end{center}

\begin{figure}[h]
 \centering
  \includegraphics[width=8cm]{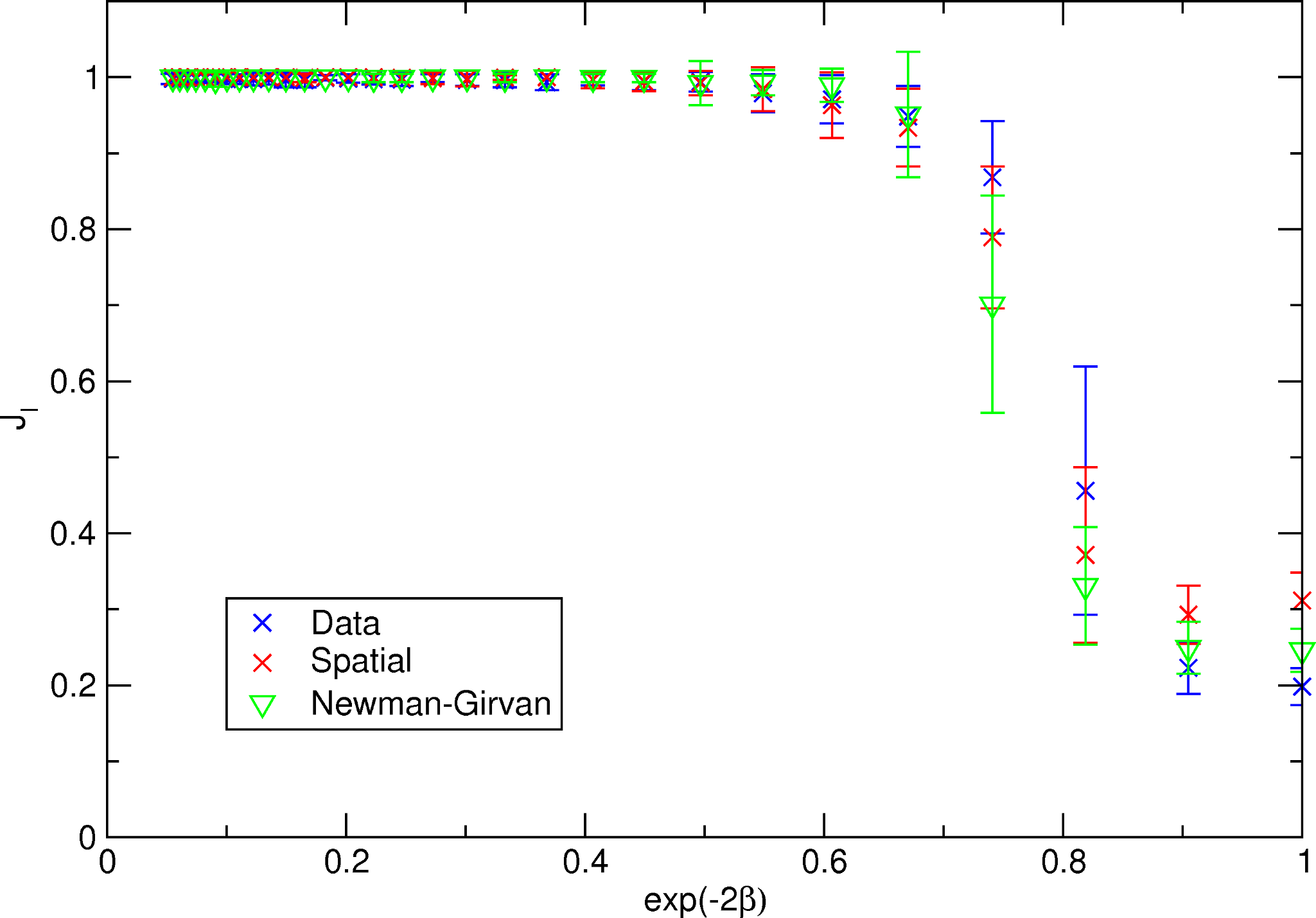} 
 \caption{Transition obtained in the case $\ell\gg L$ from the
          detectable to the undetectable community structure
          regions. This transition was described in
          \cite{Decelle:2011} for the stochastic block model which
          corresponds to our model with $q=2$ attributes when the
          effect of space is absent, i.e. $\ell$ large ($\ell=4.0$ in
          the actual simulation). The control parameter is then
          $\exp(-2\beta)$ and the Jaccard index is our order
          parameter. All the three community detection methods
          discussed in this paper display the same behavior adding
          evidence to the universality of the transition presented in
          \cite{Decelle:2011}.
    }
 \label{fig:figure_6}
\end{figure}

We note that the behavior of the error bar sizes in these figures
\ref{fig:figure_4}, is interesting.  For $\beta \ell \ll 1$ and
$\beta\ell \gg 1$, the error in the modularity estimate is relatively
small. The error bar -or equivalently the fluctuations of the Jaccard
index- are the largest for $\beta\ell\simeq 1$. In this region, the
community detection methods are thus more sensitive to small
fluctuations of the network which implies a peak in the
`susceptibility' of the system. This behavior is reminiscent of the
phase transition between detectability and non-detectability presented
in \cite{Nussinov:2011,Decelle:2011}. Indeed, in figure \ref{fig:figure_6} we show the
limiting case of $l\gg L$ (here we choose numerically $l=4$ and $L=1$)
for which the effect of space is irrelevant. In this limit, our model
becomes equivalent to the stochastic block model of
\cite{Decelle:2011} with $q=2$ possible values of the attribute. In
our case the control parameter ($c_{out}/c_{in}$ in
\cite{Decelle:2011}) is $\exp(-2\beta)$, while the order
parameter is the Jaccard index. It is clear from Fig. \ref{fig:figure_6}
that the same effect is present (see figure 2 in \cite{Decelle:2011})
even if the critical point is shifted due to a different community
detection method and another definition of the order parameter.
Moreover, respect to the result in \cite{Decelle:2011}, in the
undetactable regime ($\beta=0$), the value of the order parameter is
not zero. As mentioned above, for a completely random partition the
$J_I$ is $J_I=0.08\pm 0.05$.  We observe that in our case we are a
little bit above because it is known that even for a random network
the modularity can be positive \cite{Amaral:2004}
and in this way the maximization of the
modularity extracts a subset of the ensemble of all the possible
partitions that increases the average modularity and consequently the
average Jaccard index.

We thus recover the results of \cite{Decelle:2011} and in addition our result seems to
point to the existence of a spatial phase transition actually independent of
the community detection method used.

\begin{figure}[h]
 \centering
  \includegraphics[width=8cm]{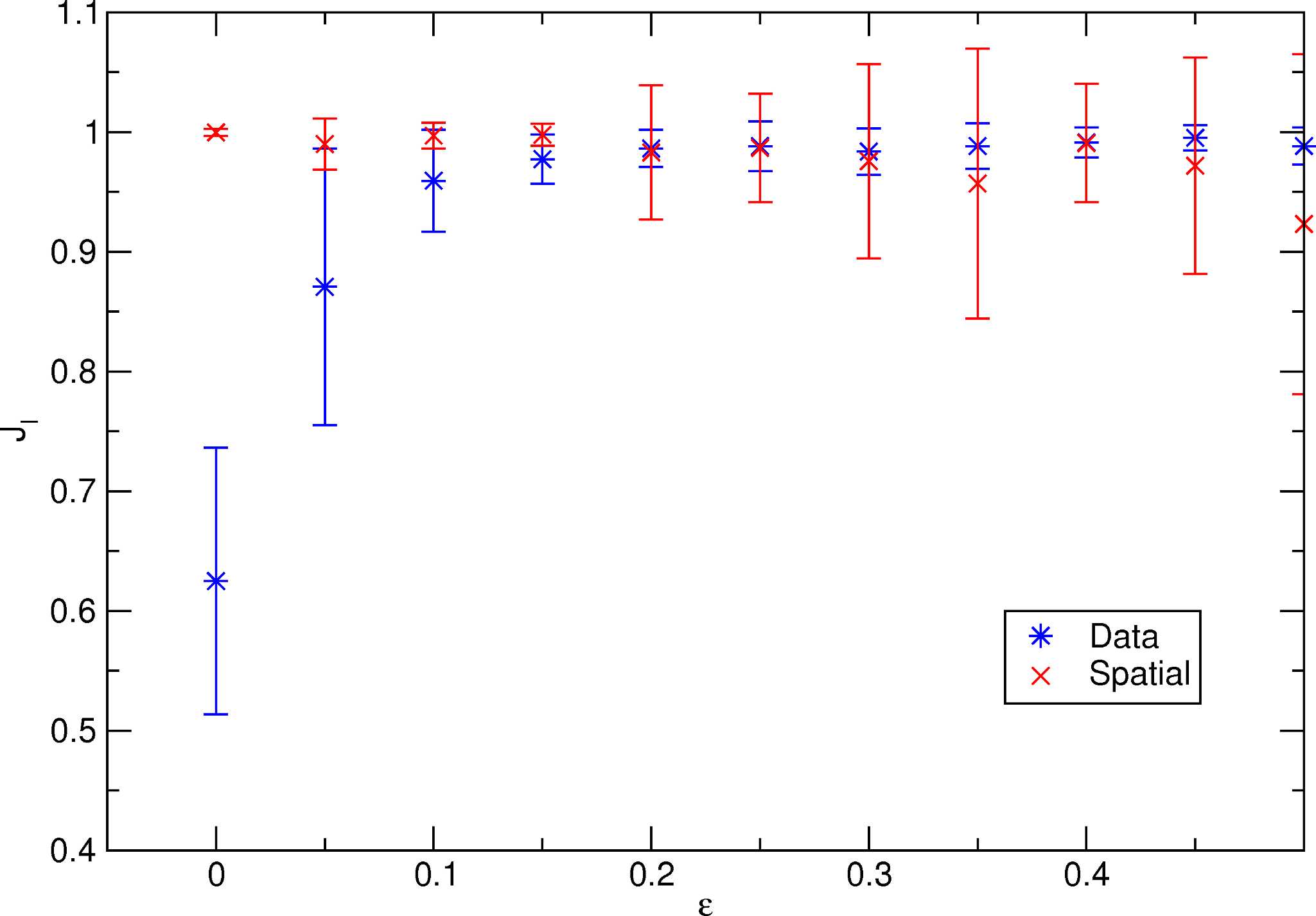} 
 \caption{Performances of the Spatial and Data modularity
  formulations. We show here the case $\beta \ell=0.8$ where there is a
  crossover in the performances around $\epsilon\simeq 0.25$. Below
  this value $\epsilon=0.25$ the Spatial method performs better and
  above the Data method is slightly better.
    }
 \label{fig:figure_7}
\end{figure}

Finally, we checked the performances of the Data and Spatial
formulations looking at the $J_I$ values when varying the $\epsilon$
parameter for a fixed $\beta \ell$ value (see
Fig.~\ref{fig:figure_7}). For each value of $\epsilon$ an higher $J_I$
value signals a better behavior since it is closer to the maximum value
$J_I=1$.  We choose first the value $\beta \ell=0.8$ (we also tested $\beta
\ell=1.0$ which gives similar results). There is a crossover in the performances
around $\epsilon\simeq 0.25$. Below this value, the Spatial method performs better
 while above that point the Data method does slightly better. This result
 thus shows that there can be a non-negligible range of correlations (measured
 here by $\epsilon$) for which the spatial community detection results can be
 incorrect.

\section{Discussion}

In this paper we propose a simple model which allows us to test
community detection on spatial networks.  {\color{black} Our model
  generates simple graphs that mix both geographical properties and
  attributes. In the literature many other spatial network models have
  been introduced for which nodes are connected each other through a
  certain spatial rule.  Examples range from the growth of street
  networks to the evolution of the territorial infrastructural
  networks (see \cite{Barthelemy:2011} for an extensive list of this
  kind of models).  Moreover a whole class of models that study node
  properties and their aggregation has recently been introduced and
  one of the most important of them is the stochastic block model in
  which a combination of various kind of node attributes are present.
  The novelty of our approach is to study at the same time these
  various aspects (geography and attributes), and, up to our knowledge, our
  model is the first one that considers simultaneously the two
  factors, space and attributes, in the context of community
  detection.}

In particular, we explicitly show that the existence of correlations
between attributes and space drastically affects the result of
community detection. The results presented in this study show that
community detection in spatial networks should be taken with great
care, and that including space in community detection methods could
lead to results difficult to interpret. We show that for weak
correlations, most community detection methods work, but that for
stronger correlation community detection methods which remove the
spatial component of the network can lead to incorrect results. It is
thus important to have some information on the correlations between
space and attributes in order to assess the validity of the results of
community detection methods. In practical applications however, these
attributes-space correlations are generally not known and this calls
for the need of new approaches, for example such as community
detection methods including in some tunable form the existence of such
correlations.

\begin{acknowledgments}

We thank Gianni Mula for enlightening conversations. MB thanks
Linkalab and the Department of Physics of the University of Cagliari
for their warm hospitality during the early stage of this work.
VDL acknowledges the support of the operating program of Regione Sardegna 
(European Social Fund 2007-2013), L.R.7/2007, 
“Promotion of scientific research and technological innovation in Sardinia”.
\end{acknowledgments}



\bibliographystyle{unsrt}
\bibliography{biblio}

\end{document}